\documentclass[preprint,authoryear,12pt]{elsarticle}

\usepackage{graphicx}
\usepackage{stfloats}
\usepackage{subcaption}
\usepackage{tikz}

\usepackage{amssymb}
\usepackage{amsthm}
\usepackage{amsmath} 

\usepackage{verbatim}
\usepackage{cite}
\usepackage{epstopdf}
\usepackage{acronym}
\usepackage{ifthen}
\usepackage{multirow}
\usepackage{algorithm, algpseudocode}

\journal{Computational and Applied Mathematics}

\begin{document}


\newtheorem*{quest}{Question}
\newtheorem{df}{Definition}
\newtheorem{eg}{Example}
\newtheorem{cor}{Corollary}
\newtheorem{thm}{Theorem}
\newtheorem{conj}{Conjecture}
\newtheorem{lem}{Lemma}

\newenvironment{ans}{\par\indent \emph{Ans:}}{\noindent \qedsymbol}    

\acrodef{VA}{Variable Annuity}
\acrodefplural{VA}{Variable Annuities}
\acrodef{MC}{Monte Carlo}
\acrodef{IDW} {Inverse Distance Weighting}
\acrodef {RBF} {Radial Basis Function}
\acrodef {AV} {Account Value} 
\acrodef {GD} {Guaranteed Deat Benefit Value} 
\acrodef {CDF} {Cumulative Distribution Function}
\acrodef {LHS} {Latin Hypercube Sampling}
\acrodef {LSMC} {Least Squares Monte Carlo}
\acrodef {SCR} {Solvency Capital Requirement}
\acrodef {MSE} {Mean Squared Error}
\acrodef {NAG} {Nestrov's Accelerated Gradient}
\acrodef {RSS} {Replicated Stratified Sampling}
\acrodef {CEIOP}{Committee of European Insurnace and Occupational Pensions Supervisors}
\acrodef {AC}{Available Capital}
\acrodef {MVA}{Market Value of Assets} 
\acrodef {MVL}{Market Value of Liabilities}
\acrodef {SE}{Social Economy}
\acrodef {MLE}{Maximum Likelihood Estimation}

\def\bibsection{\section*{References}}

\begin{frontmatter}



\title{Efficient Valuation of SCR via a Neural Network Approach}


\author[CSUT]{Seyed Amir Hejazi}
\ead{amir@cs.toronto.edu}
\author[CSUT]{Kenneth R. Jackson}
\ead{krj@cs.toronto.edu}
\address[CSUT]{Department of Computer Science, University of Toronto, Toronto, ON, M5S 3G4, Canada}

\begin{abstract}
As part of the new regulatory framework of Solvency II, introduced by the European Union, 
insurance companies are required to monitor their solvency by computing a key risk 
metric called the Solvency Capital Requirement (SCR). The official description of the SCR 
is not rigorous and has lead researchers to develop their own mathematical frameworks for calculation of 
the SCR. These frameworks are complex and are difficult to implement. Recently, Bauer et al.\ 
suggested a nested Monte Carlo (MC) simulation framework to calculate the 
SCR. But the proposed MC framework is computationally expensive even for a simple insurance 
product. In this paper, we propose incorporating a neural network approach into the nested simulation 
framework to significantly reduce the computational complexity in the calculation. 
We study the performance of our neural network approach in estimating the SCR for 
a large portfolio of an important class of insurance products called Variable Annuities (VAs). 
Our experiments show that the proposed neural network approach is both efficient and 
accurate.

\end{abstract}


\end{frontmatter}



\section{Introduction}\label{sec:intro}
The Solvency II Directive is the new insurance regulatory framework within the European Union. Solvency II 
enhances consumer protection by requiring insurers to monitor the risks facing their organization. An 
integral part of Solvency II is the \ac{SCR} that reduces the risk of insurers' insolvency. 
\ac{SCR} is the amount of reserves that an insurance company must hold to cover any losses within a one 
year period with a confidence level of $99.5\%$. 

The calculation standards are described in the 
documents of the \ac{CEIOP} (e.g., \citep{CEIOP11}). The regulation allows insurance companies to use either 
the standard formula or to develop an internal model based on a market-consistent valuation of assets and 
liabilities. Because of the imprecise language of the aforementioned standards, many insurance companies 
are struggling to implement the underlying model and to develop efficient techniques to do the necessary calculations. 
In \citep{Christiansen14, Bauer12}, rigorous mathematical definitions of \ac{SCR} are provided. 
Moreover, \citep{Bauer12} describes an implementation of a simplified, but approximately equivalent, 
notion of \ac{SCR} using nested \ac{MC} simulations. 

The results of the numerical experiments in \citep{Bauer12} to find the \ac{SCR} for a simple insurance product 
show that the proposed nested \ac{MC} simulations are too expensive, even for their simplified notion of \ac{SCR}.  
Hence, insurance companies cannot directly use the proposed \ac{MC} approach to find the \ac{SCR} for their large 
portfolios of insurance products. In this paper, we propose a neural network approach to be used within the nested 
\ac{MC} simulation framework to ameliorate the computational 
complexity of \ac{MC} simulations which allows us to efficiently compute the \ac{SCR} for large portfolios of 
insurance products. We provide insights into the efficiency of the proposed extension of the \ac{MC} simulation framework 
by studying its performance in computing the \ac{SCR} for a large portfolio of \acp{VA}, a well-known and important 
class of insurance products. 

A \ac{VA} is a tax-deferred retirement vehicle that allows a 
policyholder to invest in financial markets by making payment(s) into a predefined set of sub-accounts set 
up by an insurance company. The investment of the policyholder should be payed back as a lump-sum payment or a series 
of contractually agreed upon payments over a period of time in the future. \ac{VA} products provide embedded 
guarantees that protect the investment of a policyholder in a bear market and/or from mortality risk \citep{Geneva13}. 
For a detailed description of \ac{VA} products and the different types of guarantees offered in these products, 
see our earlier paper \citep{Hejazi15} and the references therein.

Because of the innovative structure of embedded guarantees in \ac{VA} products, insurance companies 
have been successful in selling large volumes of these products \citep{IRI11}. As a result, \ac{VA} 
products are a large portion of the investment market around the globe and big insurance companies have 
accumulated large portfolios of these products. The embedded guarantees of \ac{VA} products expose 
insurers to a substantial amount of market risk, mortality risk, and behavioral risk. Hence, big insurance 
companies have developed risk management programs to hedge their exposures, especially after the market crash 
of 2008. 

The rest of this paper is organized as follows. In Section \ref{sec:scr_def}, we describe the mathematical definition 
of \ac{SCR} as well as its simplified, almost equivalent, version described in \citep{Bauer12}. In Section \ref{sec:ns_approach}, 
we describe a modification of the nested simulation approach of \citep{Bauer12} that we use to approximate the \ac{SCR}. 
Furthermore, we define a simple asset and liability structure that allows us to remove the assets from the required 
calculation of the \ac{SCR} for the portfolio. 
In Section \ref{sec:nn}, we describe the neural network framework that we use to estimate the one-year probability 
distribution of liability for the input portfolio of \ac{VA} products. 
In Section \ref{sec:ne}, we compare the efficiency and accuracy of our method to that of a simple  
nested \ac{MC} simulation approach. In Section \ref{sec:conclusion}, we conclude the paper.

\section{Solvency Capital Requirement}\label{sec:scr_def}
A rigorous treatment of \ac{SCR}\footnote{The material in this section is based largely on the discussion in \citep{Bauer12}.} 
requires the definition of \ac{AC} which is a metric that 
determines the solvency of a life insurer at each point in time. The \ac{AC} 
is the difference between the \ac{MVA} and \ac{MVL}: 

\begin{equation}\label{eq:ac}
\text{AC}_t = \text{MVA}_t - \text{MVL}_t
\end{equation}
where the subscript $t$ denotes the time, in years, at which each variable is calculated. 

Assuming the definition \eqref{eq:ac} of \ac{AC}, the \ac{SCR}, under Solvency II, is defined as the smallest 
amount of \ac{AC} that a company must currently hold to insure a non-negative \ac{AC} in one year 
with a probability of $99.5\%$. In other words, the \ac{SCR} is the smallest amount $x$ that 
satisfies the following inequality. 

\begin{equation}\label{eq:scr_c}
P(\text{AC}_1 \geq 0 | \text{AC}_0 = x) \geq 99.5\%
\end{equation}

In practice, it is hard to find the \ac{SCR} using definition \eqref{eq:scr_c}. Hence, Bauer et al.\
use a simpler, approximately equivalent notion of the \ac{SCR} which is based on the one-year loss, 
$\Delta$, evaluated at time zero: 

\begin{equation}\label{eq:delta}
\Delta = \text{AC}_0 - \frac{\text{AC}_1}{1 + r}
\end{equation}
where $r$ is the one-year risk-free rate. The \ac{SCR} is then redefined as the one-year Value-at-Risk (VaR):

\begin{equation}\label{eq:scr_a}
\text{SCR} = \text{argmin}_x \{P(\Delta > x) \leq 0.5\%\}
\end{equation}
This is the definition of the \ac{SCR} that we use in the rest of this paper. 

\section{Nested Simulation Approach}\label{sec:ns_approach}
Given the formulation of equation \eqref{eq:scr_a}, we can calculate the \ac{SCR} by first computing 
the empirical probability distribution of $\Delta$ and then computing the $99.5\%$-quantile of the 
calculated probability distribution. We can implement this scheme by the nested simulation approach of 
\citep{Bauer12}. In this section, we first outline the nested simulation approach of \citep{Bauer12} and 
then describe our modification of it to make it more computationally efficient. 

\begin{figure}[bt]
\centering
\begin{tikzpicture}
\draw[thick] (0,7) rectangle (9,6) node[midway]{Generate $N^{(p)}$ sample paths $\text{P}^{(i)}, 1 \leq i \leq N^{(p)}$};
\draw[thick,->,>=latex] (3,6) -- (3,5);
\draw[thick] (0,5) rectangle (9,4) node[midway]{Evaluate $AC_1^{(i)}, 1 \leq i \leq N^{(p)},$ for sample paths};
\draw[thick,->,>=latex] (3,4) -- (3,3);
\draw[thick] (9.5,5) rectangle (12.5,4) node[midway]{Evaluate $AC_0$};
\draw[thick,->,>=latex] (10,4) -- (3.25,3);
\draw[thick] (0,3) rectangle (9,2) node[midway]{Evaluate $\Delta^{(i)}, 1 \leq i \leq N^{(p)},$ for sample paths};
\draw[thick,->,>=latex] (3,2) -- (3,1);
\draw[thick] (0,1) rectangle (6,0) node[midway]{Sort $\Delta^{(i)}$ in ascending order};
\draw[thick,->,>=latex] (3,0) -- (3,-1);
\draw[thick] (0,-1) rectangle (13,-2) node[midway] {Output the $\lfloor N \times 0.995 + 0.5 \rfloor$ element as the approximation for \ac{SCR}};
\end{tikzpicture}
\caption{Diagram of the nested simulation approach proposed by \citep{Bauer12}. }
\label{fig:nest-sim-approach}
\end{figure}
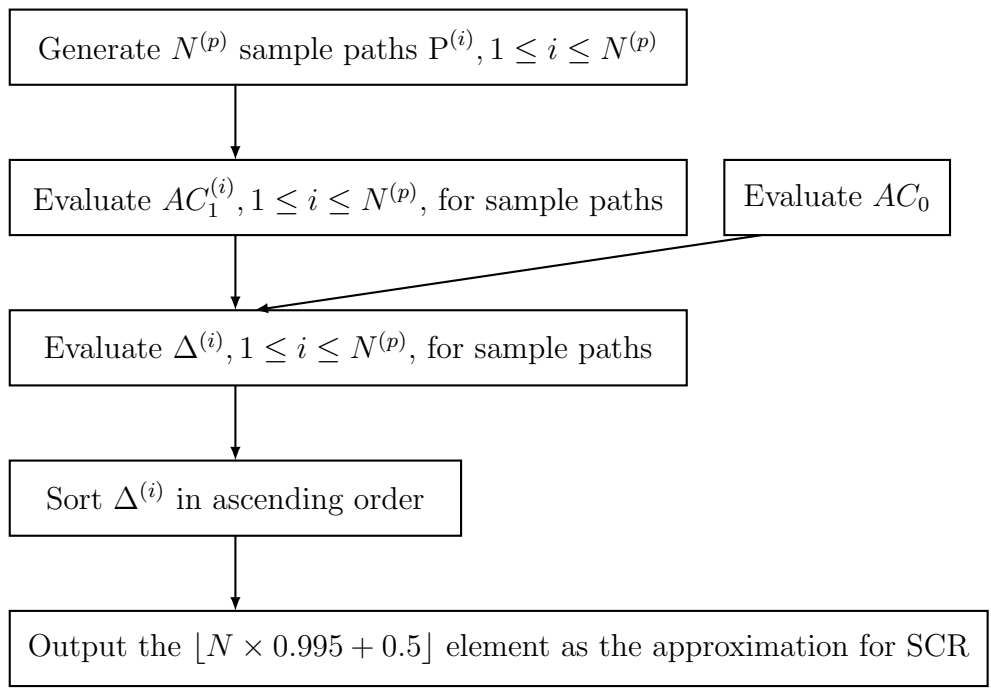

In the nested simulation approach of \citep{Bauer12}, summarized in Figure \ref{fig:nest-sim-approach}, we first 
generate $N^{(p)}$ sample paths $\text{P}^{(i)}, 1 \leq i \leq N^{(p)},$ that determine the one-year evolution of financial 
markets. Note that we are only interested in the partial state of the financial markets. In particular, we are 
only interested in the state of the financial instruments that help us evaluate the asset values and the liability 
values of our portfolio. Hence, we can generate a sample state of the financial market by drawing one sample from 
the stochastic processes that describe the value of those financial instruments of interest. 

In the nested simulation approach of \citep{Bauer12}, for each sample path $\text{P}^{(i)}$, we use a \ac{MC} simulation 
to determine the value $\text{AC}_1^{(i)}$, the available capital one year hence. We also calculate $AC_0$ via another 
\ac{MC} simulation and use that to determine the value of $\Delta^{(i)}, 1 \leq i \leq N^{(p)},$ for each sample 
path $\text{P}^{(i)}, 1 \leq i \leq N^{(p)},$ via equation \eqref{eq:delta}. The values $\Delta^{(i)}, 1\leq i \leq N^{(p)}$, 
can be used to determine the empirical distribution of $\Delta$. In order to estimate the $99.5\%$-quantile for 
$\Delta$ as required by the definition of the \ac{SCR} in equation \eqref{eq:scr_a}, we can sort the calculated 
$\Delta^{(i)}, 1\leq i \leq N^{(p)}$, values in ascending order and choose the $\lfloor N \times 0.995 + 0.5 \rfloor$ 
element amongst the sorted values as the approximation for \ac{SCR}.

The nested \ac{MC} simulation approach of Figure \ref{fig:nest-sim-approach} is computationally expensive even for simple 
insurance contracts \citep{Bauer12}. The computational complexity of the approach is caused by two factors: 
1) The value of $N^{(p)}$ can be very large \citep{Bauer12}. 2) The suggested \ac{MC} valuation of 
$\text{AC}_1^{(i)}$ for each path $\text{P}^{(i)}$ and $\text{AC}_0$, even for a single contract, is expensive and hence does not 
scale well to large portfolios of insurance products. In this paper, we focus on the latter problem and provide an 
approach to significantly reduce the cost of computing each $\text{AC}_1^{(i)}, 1\leq i \leq N^{(p)},$ and $\text{AC}_0$. 
We also briefly discuss a proposal to address the former. However, we leave a detailed development and analysis of the proposal as future work.  

To begin, we briefly outline our proposal for reducing $N^{(p)}$ to address the first problem. In the nested simulation approach suggested 
by \citep{Bauer12}, to have a good estimation of the empirical probability distribution of $\Delta$, the number of sample paths, $N^{(p)}$, 
must be large, since the $\Delta^{(i)}$ values, $1\leq i \leq N^{(p)},$ are used to approximate the probability distribution of 
$\Delta$. Consequently, many $\Delta^{(i)}$ values are needed to provide a sufficiently accurate approximation. 
Because a significant number of these values should, intuitively, be very close to each other, we suggest using a data interpolation scheme 
to reduce the cost associated with the number of sample paths, $N^{(p)}$. 
To do the interpolation, we first select/generate a small number, $N_s^{(p)}$, of paths $\text{P}_s^{(i)}, 1 \leq i \leq N_s^{(p)},$ 
and evaluate $\Delta^{(i)}$ for each path $P_s^{(i)}$. Then, we use the calculated values $\Delta_s^{(i)}, 1 \leq i \leq N_s^{(p)},$ 
of the representative paths $\text{P}_s^{(i)}, 1 \leq i \leq N_s^{(p)},$ to interpolate each $\Delta^{(i)}, 1 \leq i \leq N^{(p)},$ associated 
with each path $\text{P}^{(i)}, 1 \leq i \leq N^{(p)}$. 
The choice of the interpolation scheme that should be used depends on the distribution of the generated $N^{(p)}$ paths in the space. 
The variables that define this space are dependent on the sources of randomness in the financial 
instruments that we use to value our portfolio. In this paper, we use a simple linear interpolation scheme, described in more detail 
in Section \ref{sec:ne}, to reduce the running time of our numerical experiments. We postpone a more thorough development and analysis 
of the interpolation method to a future paper.

Now we turn to the main focus of this paper, a more efficient way to compute $\text{AC}_1^{(i)}$ for each path $\text{P}_s^{(i)}, 1 \leq i \leq N_s^{(p)}$ 
and $\text{AC}_0$. A key element in computing the $\Delta$ value via equation \eqref{eq:delta} is the calculation of \ac{AC} values. 
From \eqref{eq:ac}, we see that the calculation of \ac{AC} requires a market consistent valuation of assets and liabilities. 
Insurance companies can follow a mark-to-market approach to value their assets in a straightforward way. However, the 
innovative and complex structure of insurance products does not allow for such a straightforward calculation of liabilities. 
In practice, insurance companies often have to calculate the liabilities of insurance products 
by direct valuation of the cash flows associated with them (direct method \citep{Girard02}). Hence, 
the difficulty in calculation of \ac{SCR} is primarily associated with the difficulty in the calculation of liabilities. 

As we discuss in detail in \citep{Hejazi15}, a \ac{MC} simulation approach, as suggested in \citep{Bauer12}, 
to compute the liability of large portfolios of insurance products is very expensive. Furthermore, traditional portfolio 
valuation techniques, such as the replication portfolio approach \citep{Dembo99, Oechslin07, Daul09} and the \ac{LSMC} method 
\citep{Cathcart09, Longstaff01, Carriere96}, are not effective in reducing the computational cost. The computational complexity 
of these methods for sophisticated insurance products, such as \acp{VA}, is comparable to, or more than, the computational 
complexity of \ac{MC} schemes. Reducing the amount of computation in these methods often requires significant reduction in their accuracy. 

Recently, a spatial interpolation scheme \citep{Hejazi15, Gan13-2, Gan15} has been proposed to reduce the required computation 
of the \ac{MC} scheme by reducing the number of contracts that must be processed by the \ac{MC} method. In the 
spatial interpolation framework, we first select a sample of contracts in the space in which the insurance 
products of the input portfolio are defined. The value of interest for each sample contract is evaluated using \ac{MC} 
simulation. The outputs of the \ac{MC} simulations are then used to estimate the value of interest for other contracts in the 
input portfolio by a spatial interpolation scheme. In \citep{Hejazi15-2}, 
we describe how a neural network approach to the spatial interpolation can not only solve the problem associated with 
finding a good distance metric for the portfolio but also provide a better balance between efficiency, accuracy, 
and granularity of estimation. The numerical experiments of \citep{Hejazi15-2} provided insights into the performance 
of our proposed neural network approach in estimation of Greeks for a portfolio of \acp{VA}. We show in this paper how a similar 
neural network approach can be used within the nested \ac{MC} simulation framework to find the liabilities and subsequently 
the \ac{SCR} for an input portfolio of \ac{VA} products in an efficient and accurate manner. 

Of course, other spatial interpolation schemes could be used within the framework described in the last paragraph. However, in this paper 
we focus only on the neural network approach to show the potential of using a spatial interpolation scheme within the nested \ac{MC} 
simulation framework. Exploring the potential of other spatial interpolation schemes within this context may 
be the subject of future research. Although we are using a neural network similar to the one proposed in \citep{Hejazi15-2}, our experiments 
demonstrate that a naive usage of the neural network framework to estimate the liability values can provide no better computational 
efficiency than parallel implementation of \ac{MC} simulations. As we discuss in \citep{Hejazi15-2}, the time it takes to train the 
neural network accounts for a major part of the running time of the proposed neural network framework. Therefore, if we have to train 
the network from scratch for each realization of the financial markets ($\text{P}^{(i)}, 1 \leq i \leq N^{(p)}$) to compute  
$\Delta^{(i)}, 1\leq i \leq N^{(p)},$ the proposed neural network loses its computational superiority compared to a parallel implementation 
of the \ac{MC} simulations. To address this problem, in Section \ref{sec:ne}, we discuss a methodology to use the parameters for a 
neural network trained to compute the $\text{MVL}_1^{(i)}$ associated with $\text{P}_s^{(i)}$ as a good first guess for the parameters 
for another neural network to compute $\text{MVL}_1^{(j)}$ associated with $\text{P}_s^{(j)}$, for $i \neq j$. The proposed methodology is 
based on the idea that, for two neural networks that are trained under two market conditions that are only slightly different, 
the optimal choices of neural network parameters are likely very close to each other.

\begin{figure}[!bt]
\vspace{15mm}
\centering
\begin{tikzpicture}
\draw[thick] (2.5,14) rectangle (11.5,13) node[midway]{Generate $N^{(p)}$ sample paths $\text{P}^{(i)}, 1 \leq i \leq N^{(p)}$};
\draw[thick,->,>=latex] (7,13) -- (7,12.5);
\draw[thick] (2,12.5) rectangle (12,11.5) node[midway]{Sample/Generate $N_s^{(p)}$ sample paths $\text{P}_s^{(i)}, 1 \leq i \leq N_s^{(p)}$};
\draw[thick,->,>=latex] (4,11.5) -- (4,11);
\draw[thick,->,>=latex] (10,11.5) -- (10,11);
\draw[thick] (0,11) rectangle (7,9) node[midway]{$\begin{array}{l}\text{Compute } \text{MVA}_1^{(i)}, 1 \leq i \leq N_s^{(p)}, 
\\ \text{of the input portfolio for each } \text{P}_s^{(i)} \\ 
\text{using a mark-to-market approach}\end{array}$};
\draw[thick,->,>=latex] (3,9) -- (3,8.5);
\draw[thick] (7.5,11) rectangle (14,9) node[midway]{$\begin{array}{l}\text{Compute } \text{MVL}_1^{(i)}, 1 \leq i \leq N_s^{(p)}, 
\\ \text{of the input portfolio for each } \text{P}_s^{(i)} \\ 
\text{via the neural network approach}\end{array}$};
\draw[thick,->,>=latex] (10,9) -- (10,8.5);
\draw[thick] (1,8.5) rectangle (13,7.5) node[midway]{Compute $\text{AC}_1^{(i)} = \text{MVA}_1^{(i)} - \text{MVL}_1^{(i)}, 1 \leq i \leq N_s^{(p)}$};
\draw[thick] (2,7.5) -- (2,7);
\draw[thick] (0,7) rectangle (7,5.5) node[midway]{$\begin{array}{l}\text{Compute } \text{MVA}_0 \text{ of the input portfolio}
\\ \text{using a mark-to-market approach}\end{array}$};
\draw[thick,->,>=latex] (4,5.5) -- (4,5);
\draw[thick] (7.25,7) rectangle (14.25,5.5) node[midway]{$\begin{array}{l}\text{Compute } \text{MVL}_0 \text{ of the input portfolio}
\\ \text{via the neural network approach}\end{array}$};
\draw[thick,->,>=latex] (10,5.5) -- (10,5);
\draw[thick] (3.5,5) rectangle (10.5,4) node[midway]{Compute $\text{AC}_0 = \text{MVA}_0 - \text{MVL}_0$};
\draw[thick,->,>=latex] (10,4) -- (10,3.5);
\draw[thick,->,>=latex] (2,5.5) -- (2,3.5);
\draw[thick] (1,3.5) rectangle (13,2.5) node[midway]{Compute $\Delta_s^{(i)} = \text{AC}_1^{(i)} - \text{AC}_0, 1 \leq i \leq N^{(p)}_s$};
\draw[thick,->,>=latex] (7,2.5) -- (7,2);
\draw[thick] (0,2) rectangle (14,1) node[midway]{Interpolate $\Delta^{(i)}, 1 \leq i \leq N^{(p)},$ for $\text{P}^{(i)}$ using the $\Delta_s^{(i)}, 1 \leq i \leq N_s^{(p)},$ values};
\draw[thick,->,>=latex] (7,1) -- (7,0.5);
\draw[thick] (2,0.5) rectangle (12,-0.5) node[midway]{Sort the $\Delta^{(i)}, 1 \leq i \leq N^{(p)},$ values in ascending order};
\draw[thick,->,>=latex] (7,-0.5) -- (7,-1);
\draw[thick] (0.5,-1) rectangle (13.5,-2) node[midway]{Select the $\lfloor N \times 0.995 + 0.5 \rfloor$ element as the approximation for \ac{SCR}};
\end{tikzpicture}
\caption{Diagram of the proposed nested simulation approach.}
\label{fig:proposed-nest-sim-approach}
\end{figure}
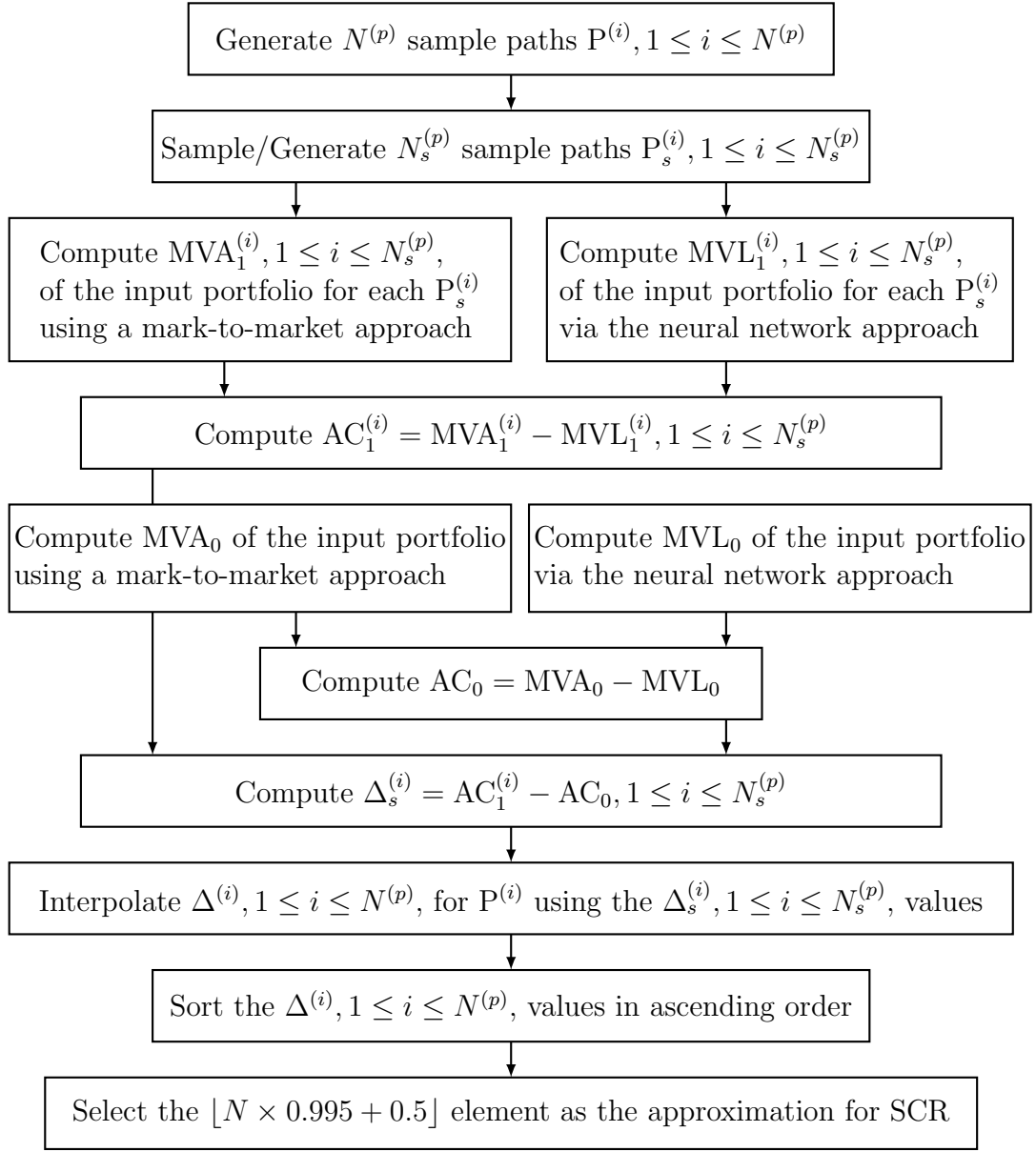

In summary, we suggest to use the extended nested simulation approach of Figure \ref{fig:proposed-nest-sim-approach} instead of the 
more standard nested simulation approach of Figure \ref{fig:nest-sim-approach} to approximate the value of \ac{SCR} via equation \eqref{eq:scr_a}.

In this paper, as mentioned earlier, our focus is on reducing the computational cost of the \ac{MC} simulations used to 
calculate the liability of large portfolios of \acp{VA}. Hence, to focus on the problem of calculating the liabilities and to make 
the analysis more tractable, we assume that the company has taken a passive approach (i.e., no hedging is involved) and the 
only asset of the company is a pool of shareholders' money $M_0$ that is invested in a money market account and hence accrues 
risk-free interest. We understand that this is a very simple, and probably unrealistic, asset structure model; however, 
using a more complex asset structure only makes the computation of asset values more time consuming and diverts our attention from 
the key issue we are focusing on in this paper which is how to improve the efficiency of the computation of the liability values. 
Moreover, note that we can use the proposed framework to calculate the portfolio liability value of the input portfolio independently 
of the evaluation of asset values. That is, a more complex model for asset valuation can be inserted into our proposed framework without 
changing our scheme for improving the efficiency of the computation of liability values. 

\pagebreak
The assets are adjusted yearly as required. The proposed simple structure of assets allows us to eliminate the assets in the definition 
of $\Delta$ in \eqref{eq:delta} as follows. 

\begin{align}\label{eq:delta_a_2}
\Delta &= \text{AC}_0 - \frac{\text{AC}_1}{1 + r} \nonumber \\
&= (M_0 - \text{MVL}_0) - (\frac{(M_0 (1 + r) - \text{MVL}_1)}{1 + r}) \nonumber \\
&= -\text{MVL}_0 + \frac{\text{MVL}_1}{ 1 + r}
\end{align} 
Hence, in our simplified problem, calculating the \ac{SCR} reduces to the problem of calculating the current liability and 
the distribution of the liability in one-year's time.

\section{Neural Network Framework}\label{sec:nn}
In this section, we provide a brief review of our proposed neural network framework. 
For a detailed treatment of this approach, in particular the reason behind our choice of network 
and training method, see our paper \citep{Hejazi15-2}. 

\subsection{The Neural Network}
Our proposed estimation scheme is an extended version of the Nadaraya-Watson kernel regression 
model \citep{Nadaraya64, Watson64}. Assuming $y(z_1), \cdots, y(z_n)$ are 
the observed values at known locations $z_1, \cdots, z_n$, our model estimates $y$ at a location 
$z$ where $y(z)$ is not known by 

\begin{equation}\label{eq:est-model}
\hat{y}(z) = \sum_{i = 1}^n \frac{G_{h_i}(z - z_i) \times y(z_i)}{\sum_{j = 1}^n G_{h_j}(z - z_j)}
\end{equation}  
where $G$ is a nonlinear differentiable function and the subscript, $h_i$,  
denotes the range of influence of each $y(z_i)$ on the estimated value. The variable $h_i$ is a location 
dependent vector that determines the range of influence of each pointwise estimator in each direction of 
feature space of the input data. In our application of interest in this paper, $y(\cdot)$ is the \ac{MC} estimation of 
liability. The variables $z_i, 1 \leq i \leq n$, are $n$ vectors in $\mathbb{R}^m$ representing the 
attributes of a sample set of $n$ representative \ac{VA} contracts in the space of the input portfolio. 
The representative \ac{VA} contracts are selected using a sampling scheme to effectively fill the space 
in which the input portfolio is defined. The size of the portfolio of representative \ac{VA} contracts 
should be much smaller than the input portfolio for the neural network scheme to be efficient. 
As we discuss in \citep{Hejazi15-2} and in more detail in \citep{Hejazi16-T}, the choice of the representative 
contracts can affect the performance, in particular the accuracy, of the neural network framework. Hence, it is 
important to use an appropriate sampling method. We postpone a discussion of the choice of an effective 
sampling method to a future paper.

We choose to implement our model \eqref{eq:est-model} using a feed-forward neural network \citep{Bishop06, Fister16} 
that allows us to fine-tune our model to find the optimum choices of the $h_i$ values that minimize our estimation error.

\begin{figure}[bt]
\centering
\begin{tikzpicture}
 \foreach \y in {2,4,8}
  \draw[red, thick] (2.25,\y - 0.75) rectangle (2.75, \y + 0.75);
 \foreach \y in {5.5,6,6.5}
  \fill[black] (2.5, \y) circle(0.05cm);
  
 \foreach \y in {2,4,8}
  \draw[black, thick] (4,\y) circle (0.25cm);
 \foreach \y in {5.5,6,6.5}
  \fill[black] (4, \y) circle(0.05cm);

 \foreach \y in {2,4,8}
  \draw[blue, thick] (6,\y) circle (0.25cm);
 \foreach \y in {5.5,6,6.5}
  \fill[black] (6, \y) circle(0.05cm);

 \foreach \y in {2,4,8}
  \foreach \yy in {2,4,8}
  {
   \draw[black,->,>=latex] (2.75,\y - 0.4) -- (3.75, \y - 0.1);
   \draw[black,->,>=latex] (2.75,\y) -- (3.75, \y);
   \draw[black,->,>=latex] (2.75,\y + 0.4) -- (3.75, \y + 0.1);
  }

 \foreach \y in {2,4,8}
  \foreach \yy in {2,4,8}
   \draw[black,->,>=latex] (4.25,\y) -- (5.75, \yy);

 \draw[black, thick] (0.5,1.75) -- (0.25,1.75) -- (0.25,8.25) -- (0.5,8.25);
 \node[rotate=90] at (0, 5){Input Layer};
 \node[] at (1.25, 2){\begin{tabular}{c}$z_1$\\ Features\end{tabular}};
 \node[] at (1.25, 4){\begin{tabular}{c}$z_2$\\ Features\end{tabular}};
 \node[] at (1.25, 8){\begin{tabular}{c}$z_n$\\ Features\end{tabular}};

 \draw[black, thick] (3,9) -- (3,9.25) -- (5,9.25) -- (5,9);
 \node at (4,9.5){Hidden Layer};
 \draw[black, thick] (6.25,8.25) -- (6.5, 8.25) -- (6.5,1.75) -- (6.25,1.75);
 \node[rotate=90] at (6.75, 5){Output Layer};
\end{tikzpicture}
\caption{Diagram of the proposed neural network. Each circle represents a neuron. Each rectangle represent the 
set of neurons that contains input features corresponding to a representative contract.}
\label{fig:nn}
\end{figure}
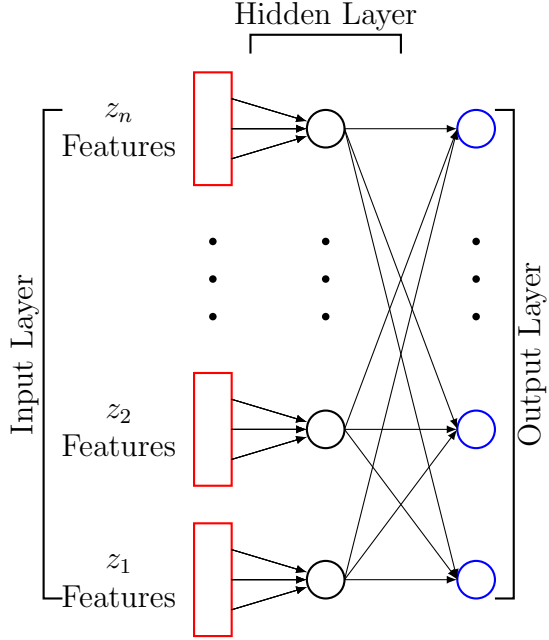

As shown in Figure \ref{fig:nn}, our feed-forward neural network is a collection 
of interconnected processing units, called neurons, which are organized in three layers. 
The first and the last layers are, respectively, called the input layer and the output layer. 
The intermediate layer is called the hidden layer.

The neurons in the first layer provide the network with the feature vector (input values).
Each neuron in the input layer represents a value in the set $\{F^c, F^-, F^+\}$. 
Each $f$ in $F^c$ has the form 

\begin{equation}
 f = \Big\{\begin{matrix} 0 & \text{if }x_c = x_{c_i}\\ 1 & \text{if }x_c \neq x_{c_i}\end{matrix}
\end{equation}
where $x_c$ represents the category of categorical attribute $c$ for input \ac{VA} policy $z$, 
and $x_{c_i}$ represents the category of categorical attribute $c$ for representative \ac{VA} policy 
$z_i$ in the sample.
Each $f$ in $F^-$ has the form $f = [t(\mathbf{x}_{n_i}) - t(\mathbf{x}_n)]^+/R_{t}$, and each 
$f$ in $F^+$ has the form $f = [t(\mathbf{x}_n) - t(\mathbf{x}_{n_i})]^+/R_{t}$. In both of these 
formulas, $\mathbf{x}_n$ is the vector containing the numeric attributes of input \ac{VA} policy $z$, 
$\mathbf{x}_{n_i}$ is the vector containing the numeric attributes of representative \ac{VA} policy $z_i$ 
in the sample, $t(\cdot)$ is a transformation (linear/nonlinear), determined by the expert user, 
that assumes a value in an interval of length $R_{t}$ and $[\cdot]^+ = \max(\cdot, 0)$. 
The $R_{t}$'s are used to scale the $f$'s so that each $f \in [0,1]$. It is well-known that scaling 
the variables in this way helps to increase the rate of convergence of the optimization method 
described in Section \ref{sec:nn-training} used to train the neural network.
In essence, our choice of input values allows different bandwidths 
($h_i$ values in \eqref{eq:est-model}) to be used for different attributes of \ac{VA} policies and in 
different directions around a representative \ac{VA} contract in the sample.

Since we are interested in calibrating the $G$ functions of equation \eqref{eq:est-model}, the 
number of neurons in the output and hidden layer are equal to the number of representative 
contracts in the sample. The inputs of neuron $i$ in the hidden layer are those values of $f$ 
in the input layer that are related 
to the representative \ac{VA} policy $i$. In other words, the input values of neuron $i$ in the hidden 
layer determine the per attribute difference of the input \ac{VA} contract $z$ with the representative 
\ac{VA} contract $z_i$ using the features $f \in \{F^c, F^-, F^+\}$.
Assuming $x_1, \cdots, x_n$ are the inputs of neuron $j$ at the hidden level, first a linear 
combination of input variables is constructed  

\begin{equation}\label{eq:activation}
a_j = \sum_{i = 1}^{n} w_{ij}x_i + b_j
\end{equation}
where parameters $w_{ij}$ are referred to as weights and parameter $b_j$ is  called 
the bias. The quantity $a_j$ is known as the activation of neuron $j$. 
The activation $a_j$ is then transformed using an exponential function to form the output of neuron $j$. 

The output of neuron $i$ in the output layer is the normalized version of the output for  
neuron $i$ in the hidden layer. Hence the outputs of the network, i.e., $o_i, i \in \{1,\cdots,n\}$, represent a 
softmax of activations in the hidden layer. These outputs can be used to estimate the value of the 
liability for input \ac{VA} $z$ as $\hat{y}(z) = \sum_{i = 1}^n o_i \times y(z_i)$, in which $y(z_i)$ is the 
value of the liability for representative \ac{VA} policy $z_i$. In summary, our proposed neural network allows 
us to rewrite equation \eqref{eq:est-model} as 

\begin{equation}\label{eq:nn-model}
\hat{y}(z) = \sum_{i = 1}^n \frac{\exp(\mathbf{\mathbf{w}_i}^T \mathbf{f_i}(z) + b_i) \times y(z_i)}{\sum_{j = 1}^n \exp(\mathbf{\mathbf{w}_j }^T \mathbf{f_j}(z) + b_j)}
\end{equation}
where vector $\mathbf{f_i}$ represents the features in the input layer that are related to the representative 
\ac{VA} policy $z_i$, and vector $\mathbf{w_i}$ contains the weights associated with each feature in 
$\mathbf{f_i}$ at neuron $i$ of the hidden layer. 

\subsection{Network Training Methodology}\label{sec:nn-training}
In order to calibrate (train) the network and find the optimal values of weights and bias parameters, 
we select a small set of \ac{VA} policies, which we call the training portfolio, as the training data for the network.  
The objective of the calibration process is to find a set of weights and bias parameters that minimizes the 
\ac{MSE} in estimation of liability values of the training portfolio. In other words, our objective 
function is

\begin{equation}\label{eq:mse-error}
E(\mathbf{w}, \mathbf{b}) = \frac{1}{2n} \sum_{k = 1}^n || \hat{y}(z_k, \mathbf{w}, \mathbf{b}) - y(z_k)||^2
\end{equation}

We use an iterative gradient descent scheme \citep{Boyd04} to train the network. However, to speed up the training process, 
we do mini-batch training \citep{Murphy12} with \ac{NAG} method \citep{Nesterov83}. In mini-batch training, in each iteration, 
we select a small number of training \ac{VA} policies at random and compute the gradient of the following error function for this batch.  

\begin{equation}\label{eq:mse-mini}
E(\mathbf{w^{(t)}}, \mathbf{b^{(t)}}) = \frac{1}{2|B^{(t)}|} \sum_{k \in B^{(t)}} || \hat{y}(z_k, \mathbf{w^{(t)}}, \mathbf{b^{(t)}}) - y(z_k)||^2
\end{equation}
where $B^{(t)}$ is the set of indices for the selected \ac{VA} policies and superscript $t$ denotes the iteration number.
Instead of updating the weights and biases by the gradient of \eqref{eq:mse-mini}, in the \ac{NAG} method, we use 
a velocity vector that increases in value in the direction of persistent reduction in the objective error 
function across iterations. We use a particular implementation of the \ac{NAG} method described in \citep{Sutskever13}. 
In this implementation of \ac{NAG}, weights and biases are updated according to the rules 

\begin{align}\label{eq:momentum}
&v_{t+1} = \mu_t v_t - \epsilon \nabla E([\mathbf{w}^{(t)}, \mathbf{b}^{(t)}] + \mu_t v_t) \nonumber \\
&[\mathbf{w}^{(t + 1)}, \mathbf{b}^{(t + 1)}] = [\mathbf{w}^{(t)}, \mathbf{b}^{(t)}] + v_{t + 1}
\end{align}
where $v_t$ is the velocity vector, $\mu_t \in [0, 1]$ is known as the momentum coefficient and $\epsilon$ is the 
learning rate. The momentum coefficient is an adaptive parameter defined by 

\begin{equation}\label{eq:momentum-coeff}
\mu_t = \min (1 - 2^{-1 - \log_2(\lfloor \frac{t}{50} \rfloor + 1)}, \mu_{\max})
\end{equation}  
where $\mu_{\max} \in [0, 1]$ is a user defined constant.

Because of the amount of investments and the structure of guarantees in \ac{VA} products, the liability values 
can become large. Big liability values can result in big gradient values which produce big jumps in the updates 
of \eqref{eq:momentum}. Therefore, to avoid numerical instability, only in the training stage, we normalize the values of 
$y(z_i)$ in \eqref{eq:mse-mini}, by dividing each $y(\cdot)$ value by the range of guarantee values in the input portfolio. 

If we allow the network to train for a long enough time, it will start to converge towards a local optimum. 
Depending on our choice of the representative contracts and the training data, further training of the network 
after a certain number of iterations might result in overfitting or might not result in significant change in the value 
of weights and biases, and the associated error. To avoid these pitfalls, we  
use a set of randomly selected \ac{VA} policies from the input portfolio as our validation portfolio \citep{Murphy12} and 
stop the training using a two step verification process. First, we observe if the \ac{MSE} of the training data 
drops dramatically or if there is an initial decrease in the \ac{MSE} of the validation portfolio to a local minimum followed by an 
increase in the \ac{MSE} of the validation portfolio. Once any of these events, called stopping events, happens, we train the 
network for a few more iterations until the mean of the network's liability estimates for the validation portfolio is within 
a $\delta$ relative distance of the mean of the \ac{MC} estimated liability of the validation portfolio via \ac{MC} simulations or a maximum number 
of training iterations is reached. The relative distance between the network estimated liablity $L_{NN}$ and the \ac{MC} estimated 
liability $L_{MC}$ is calculated as 

\begin{equation}
 \text{dist} = \left|\frac{L_{NN} - L_{MC}}{L_{MC}}\right|
\end{equation}

\begin{figure}[bt]
\centering
\includegraphics[width=\textwidth, height=0.5\textheight]{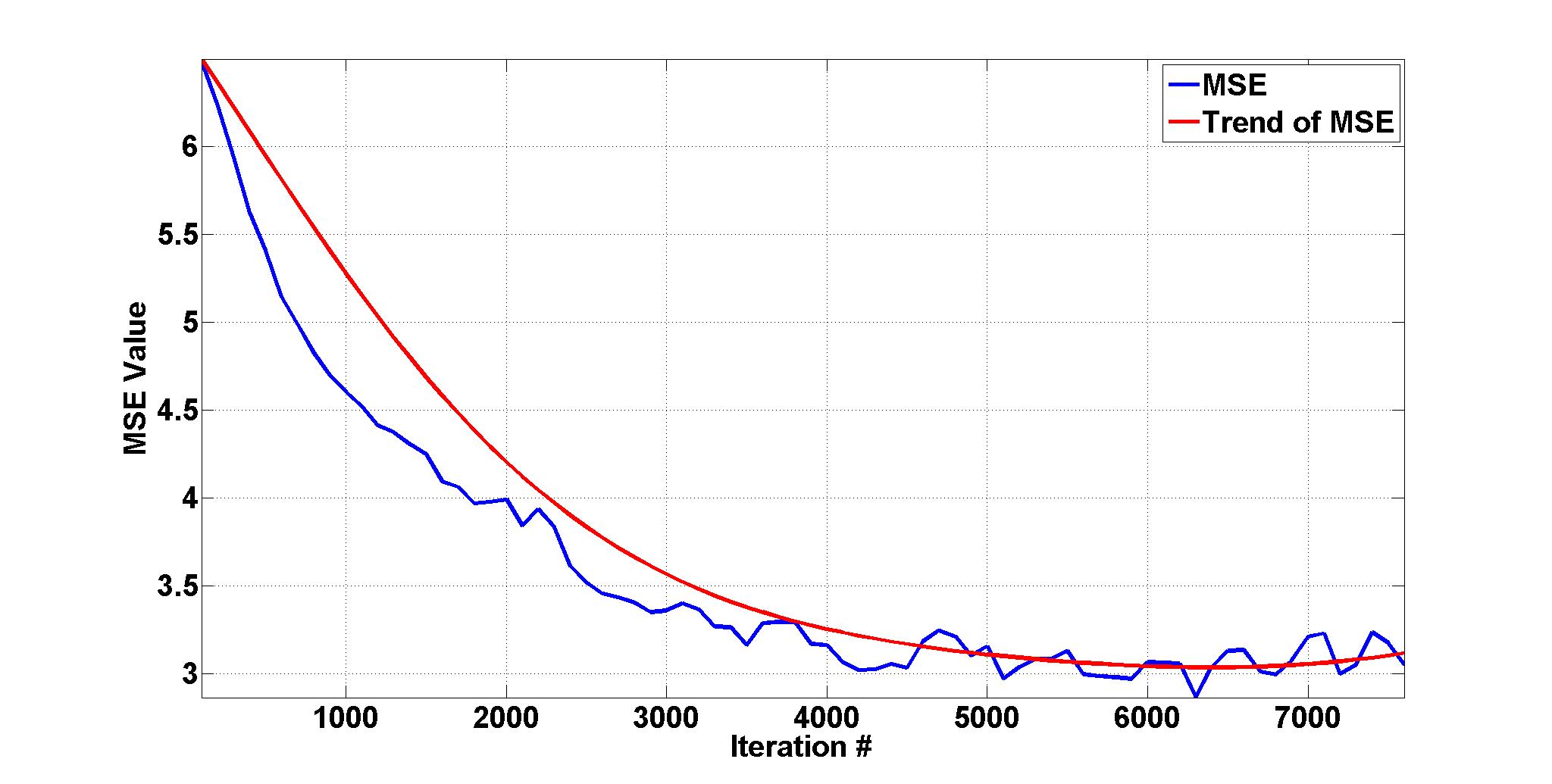}
\caption{\ac{MSE} of the validation set and the trend in the \ac{MSE} as a function of the iteration number for a run of the training algorithm. 
The trend is found using a moving average with a window size of 10 and then polynomial fitting with a polynomial of degree 6.}
\label{fig:validation-err}
\end{figure}

As shown in the graph of Figure \ref{fig:validation-err}, the actual graph of the \ac{MSE} for the validation portfolio or the 
training portfolio as a function of iteration number might be volatile. However, a general trend exists in the data. To 
make the trend clearer, we use a simple moving average with a window of $\bar{W}$ to smooth the data and polynomial 
fitting of the smoothed data. We detect stopping events using a window of length $W$ on the polynomial approximation of 
the \ac{MSE} values. A stopping event occurs if the \ac{MSE} of the validation set 
has increased in the past $W - 1$ recorded values after attaining a minimum.  We evaluate the \ac{MSE} values 
of the validation set every $I^{th}$ iteration of the training, to avoid slowing down the training process. 
$I$, $W$ and $\bar{W}$ are user defined parameters and are application dependent.

The neural network outlined above is discussed in more detail in \citep{Hejazi15-2}, where we also discuss how to choose the free 
parameters described above.

\section{Numerical Experiments}\label{sec:ne}
In this section, we demonstrate the effectiveness of the proposed neural 
network framework in calculating portfolio liability values in the proposed 
nested simulation approach of Section \ref{sec:ns_approach}. To do so, we estimate 
the \ac{SCR} for a synthetic portfolio of $100,000$ \ac{VA} contracts assuming the 
financial structure of assets as described in Section \ref{sec:ns_approach} that 
allows us to use equation \eqref{eq:delta_a_2}.

\begin{table}[!bt]
 \centering
 \begin{tabular}{|l|l|}
  \hline
  {\bf Attribute} & {\bf Value}\\ 
  \hline
  Guarantee Type & \{GMDB, GMDB + GMWB\} \\
  \hline
  Gender & \{Male, Female\}\\
  \hline
  Age & $\{20, 21, \ldots, 60\}$\\
  \hline
  Account Value & $[1e4, 5e5]$ \\ 
  \hline 
  Guarantee Value & $[0.5e4, 6e5]$ \\
  \hline
  Widthrawal Rate & $\{0.04, 0.05, 0.06, 0.07, 0.08\}$\\
  \hline
  Maturity & $\{10, 11, \ldots, 25\}$\\
  \hline
 \end{tabular}
 \caption{GMDB and GMWB attributes and their respective ranges of values.}
 \label{tb:portfolio}
\end{table}

Each contract in the portfolio is assigned attribute 
values uniformly at random from the space defined in Table \ref{tb:portfolio}. 
The guarantee values (death benefit and withdrawal benefit) of GMWB riders are 
chosen to be equal\footnote{This is typical of the beginning of 
the withdrawal phase.}, but they are different than the account value. 
The account values of the contracts follow a simple log-normal 
distribution model \citep{Hull06} with a risk free rate of return of $\mu = 3\%$, 
and volatility of $\sigma = 20\%$. 

We acknowledge that this model of account value is very simple; we use it here 
to make our computations more tractable. A more complex model increases the number of \ac{MC} 
simulations that is required to find liability values and affects the distribution 
of one-year-time's liability values. A more complex model of account values may 
also necessitate the use of more complicated valuation techniques for which the 
computational complexity is much more than simple \ac{MC} simulations. However, regardless 
of the changes imposed by using a more complex model to describe the dynamics of
the account value, the proposed nested simulation framework incorporating a neural 
network approach, described in Sections \ref{sec:ns_approach} and \ref{sec:nn}, 
can be used to calculate the \ac{SCR}. Therefore, to focus on our neural network 
approach in this paper, we have chosen to use a simple account value model to avoid 
distracting the reader with a lengthy description of a more complex account value model. 

A change in the probability distribution of one-year-time's liability values  
changes the probability distribution of $\Delta^{(i)}, 1 \leq i \leq N^{(p)},$ values. 
A change in the probability distribution of $\Delta^{(i)}, 1 \leq i \leq N^{(p)},$ values  
can increase the number of sample $\text{P}_s^{(i)}, 1 \leq i\leq N_s^{(p)},$ paths 
and affect the choice of the interpolation scheme that should be used to calculate the $\Delta^{(i)}, 1 \leq i \leq N^{(p)},$ 
values. As mentioned earlier, in this paper, our focus is not on the choice of 
the interpolation scheme and/or the size $N_s^{(p)}$ of the sample paths that should be used here. 
We leave a more thorough study of these questions to future work. We can still repeat the experiments 
of this section and arrive at the same conclusions even if we directly compute the $\Delta^{(i)}, 1 \leq i \leq N^{(p)},$ 
values for the original $N^{(p)}$ paths $\text{P}^{(i)}, 1 \leq i \leq N^{(p)}$; however, 
the running times will be much bigger.

We can increase the number of \ac{MC} simulations or use more complex techniques to 
value liability values. Both of these approaches only slightly increase the running time of our 
proposed neural network approach to calculate liability values and hence only slightly increase the 
running time of our proposed nested simulation approach of Section \ref{sec:ns_approach}. However 
the aforementioned approaches significantly increase the running time of the nested simulation approach 
of \citep{Bauer12}. The nested simulation approach of \citep{Bauer12} evaluates the liability values 
for each \ac{VA} in the input portfolio; however, our proposed neural network framework only evaluates 
the liability values for the selected number of sample \ac{VA} contracts of the representative portfolio, 
training portfolio, and the validation portfolio and then does a spatial interpolation to find the 
liability values for the \ac{VA}s in the input portfolio. Increasing the time to calculate the per 
\ac{VA} liability value linearly increases the running time in the nested simulation approach of 
\citep{Bauer12}. The size of the representative portfolio, the training portfolio and the validation 
portfolio combined in practice is much smaller than the size of the input portfolio. Therefore, 
increasing the time to calculate per \ac{VA} liability only affects the total running time of the 
neural network to the extend that the calculation of liability values for the representative portfolio, 
the training portfolio, and the validation portfolio can affect the training time which in most of our 
experiments is not significant. Most of the computing time for the neural network approach is consumed 
in training the network. 

We use the framework of \citep{Gan15} to value each \ac{VA} contract. 
Similar to \citep{Hejazi15}, we use $10,000$ \ac{MC} simulations to value each contract. 
In our experiments, we use the mortality rates of the 1996 IAM mortality tables provided by 
the Society of Actuaries.  

We implement our experiments in Java and run them on a machine with dual quad-core Intel X5355 CPUs. 
For each valuation of the input portfolio using the \ac{MC} simulations, we divide the input portfolio 
into 10 sub-portfolios, each with an equal number of contracts, and run each sub-portfolio on one thread, 
i.e., a total of $10$ threads, to value these $10$ sub-portfolios in parallel. We use a similar parallel 
processing approach to value the representative contracts, the training portfolio and the validation portfolio. 
Although we use the parallel processing capability of our machine for \ac{MC} simulations, we do not use 
parallel processing to implement our code for our proposed neural network scheme: our neural network code is 
implemented to run sequentially on one core. However, there is significant potential for parallelism in 
our neural network approach, which should enable it to run much faster. We plan to investigate this in a future paper.

\subsection{Network Setup}\label{sec:net-setup}
Although a sagacious sampling scheme can significantly improve the performance of the network (see \citep{Hejazi16-T}), 
for the sake of simplicity, we use a simple uniform sampling method similar to that used in \citep{Hejazi15-2}. 
We postpone the discussion on the choice of a better sampling method to future work. 
We construct a portfolio of all combinations of attribute values defined in Table \ref{tb:rep_contracts1}. 
In each experiment, we randomly select $300$ \ac{VA} contracts from the aforementioned portfolio 
as the set of representative contracts. 

\begin{table}[!bt]
 \centering
 \begin{tabular}{|l|l|}
  \hline
  & Experiment 1\\
  \hline
  Guarantee Type & \{GMDB, GMDB + GMWB\}\\ 
  \hline
  Gender & \{Male, Female\}\\
  \hline
  Age & $\{20, 30, 40, 50, 60\}$\\
  \hline 
  Account Value & $\{1e4, 1e5, 2e5, 3e5, 4e5, 5e5\}$\\
  \hline
  Guarantee Value & $\{0.5e4, 1e5, 2e5, 3e5, 4e5, 5e5, 6e5\}$\\
  \hline 
  Withdrawal Rate & $\{0.04, 0.08\}$\\ 
  \hline 
  Maturity & $\{10, 15, 20, 25\}$\\
  \hline
 \end{tabular}
 \caption{Attribute values from which representative contracts are generated for experiments.}
 \label{tb:rep_contracts1}
\end{table}

As discussed in Section \ref{sec:nn}, in addition to the set of representative contracts, we need 
to introduce two more portfolios, the training portfolio and the validation portfolio, to 
train our neural network. For each experiment, we randomly select $250$ \ac{VA} contracts 
from the input portfolio as our validation portfolio. The training portfolio, in each experiment, 
consists of $200$ contracts that are selected uniformly at random from the set of \ac{VA} contracts of all 
combinations of attributes that are presented in Table \ref{tb:training_portfolio}. 
In order to avoid unnecessary overfitting of the data, the attributes of 
Table \ref{tb:training_portfolio} are chosen to be different than the corresponding values in 
Table \ref{tb:rep_contracts1}.

\begin{table}[!bt]
 \centering
 \begin{tabular}{|l|l|}
  \hline
  & Experiment 1\\
  \hline
  Guarantee Type & \{GMDB, GMDB + GMWB\}\\ 
  \hline
  Gender & \{Male, Female\}\\
  \hline
  Age & $\{23, 27, 33, 37, 43, 47, 53, 57\}$\\
  \hline 
  Account Value & $\{0.2e5, 1.5e5, 2.5e5, 3.5e5, 4.5e5\}$\\
  \hline
  Guarantee Value & $\{0.5e5, 1.5e5, 2.5e5, 3.5e5, 4.5e5, 5.5e5\}$\\
  \hline 
  Withdrawal Rate & $\{0.05, 0.06, 0.07\}$\\ 
  \hline 
  Maturity & $\{12, 13, 17, 18, 22, 23\}$\\
  \hline
 \end{tabular}
 \caption{Attribute values from which training contracts are generated for experiments.}
 \label{tb:training_portfolio}
\end{table}

We train the network using a learning rate of $20$, a batch size of $20$ and we set $\mu_{\max}$ 
to $0.99$. Moreover, we fix the seed of the pseudo-random number generator that we use to select 
mini batches to be zero. For a given set of the representative contracts, the training portfolio, 
and the validation portfolio, fixing the seed allows us to reproduce the trained network. We set the 
initial values of the weight and bias parameters to zero.

We estimate the liability of the training portfolio and the validation portfolio every $50$ iterations 
and record the corresponding \ac{MSE} values. We smooth the recorded \ac{MSE} values using a moving average 
with a window size of $10$. Moreover, we fit a polynomial of degree $6$ to the smoothed \ac{MSE} 
values and use a window size of length $4$ to find the trend in the \ac{MSE} graphs. In the final stage of 
the training, we use a $\delta$ of $0.005$ as our threshold for maximum relative distance in estimation of 
the liabilities for the validation portfolio.

We use the rider type and the gender of the policyholder as the categorical features in $F^c$. 
The numeric features in $F^+$ are defined as follows.  

\begin{align}\label{eq:nn-numeric-feature}
&f(z, z_i) = \frac{[t(x) - t(x_i)]^{+}}{R_t} 
\end{align}
In our experiments, $t$ can assume the values maturity, age, AV, GD, GW and withdrawal rate, 
$R_t$ is the range of values that $t$ can assume, $x$ and $x_i$ are vectors denoting the 
numeric attributes of the input \ac{VA} contract $z$ and the representative contract $z_i$, respectively. 

We define the features of $F^-$ in a similar fashion by swapping $x$ and $x_i$ on the right side of 
equation \eqref{eq:nn-numeric-feature}.

\subsection{Performance}\label{sec:performance}
The experiments of this section are designed to allow us to compare the efficiency and 
the accuracy of the proposed neural network approach to the nested simulations with the 
nested \ac{MC} simulation approach of \citep{Bauer12}. 
In each experiment, we use $N^{(p)} = 40,000$ realizations of the market to estimate the 
empirical probability distribution of $\Delta$. As we describe in Section \ref{sec:ns_approach}, 
the particular simple structure of assets that we use allows use to use equation \eqref{eq:delta_a_2} to 
evaluate $\Delta$. By design, the liability value of the \ac{VA} products that we use 
in our experiments is dependent on their account values. As mentioned earlier, the account 
values follow a log-normal distribution model. Hence, we can describe the state of the financial market 
by the one-year's time output of the stochastic process of the model. 
Assuming a price of $A_0$ as the current account value of a \ac{VA}, each realization of the market 
corresponds to a coefficient $C_1$, from the above-mentioned log-normal distribution, that allows us 
to determine the account value in one year's time as $A_1 = C_1 \times A_0$. 

To come up with sample paths $\text{P}_s^{(i)}$, we determine a range (interval) based on the maximum value 
and the minimum value of the generated $40,000$ $C_1^{(i)}, 1 \leq i \leq 40,000,$ coefficients that 
describe the state of the financial markets in one-year's time and divide that range into $99$ equal length sub-intervals. 
We use the resulting $100$ end points, $C_{s_1}^{(i)}, 1 \leq i \leq 100,$ as the sample paths $\text{P}_s^{(i)}, 1 \leq i \leq 100$. 

If one graphs the resulting $\Delta_s^{(i)}, 1\leq i \leq 100,$ values as a function of the  
$C_{s_1}^{(i)}, 1 \leq i \leq 100,$ values that describe the evolution of the financial markets, 
the resulting curve is very smooth and the $100$ points are very close to each other in the space.
Because of this, we chose to interpolate the value of $\Delta^{(i)}, 1\leq i \leq 40,000$ for the 
aforementioned $40,000$ realizations of the financial markets ($C_1^{(i)}, 1\leq i \leq 40,000,$) 
by a simple piecewise-linear interpolation of the $\Delta_s^{(i)}, 1\leq i \leq 100,$ values. As we 
discuss later, the choice of a piecewise-linear interpolator might not be optimal. As noted earlier  
in Section \ref{sec:ns_approach}, we use it here as a first simple choice for an interpolation. We plan 
to study the choice of possibly more effective interpolations later.

To estimate the liability values for each of the sample paths $\text{P}_s^{(i)}$ via the proposed 
neural network framework, we first generate the representative portfolio, the training portfolio, 
and the validation portfolio. We then train the network using the liability of values at time 0 
(current liability) of \acp{VA} in these portfolios. We use the trained network to estimate the 
liability of the input portfolio at time 0. 

As we mention in Section \ref{sec:ns_approach}, if we train the network before estimating each 
liability, the running time of the proposed neural network approach, because of the significant 
time it takes to train the network, is no better than a parallel implementation of \ac{MC} simulation. 
To address this issue, we use the above-mentioned trained network for the liability values at time 0 
to estimate the one year liability of the input portfolio for each end point, $C_{s_1}^{(i)}, 1 \leq i \leq 100$. 
However, before each estimation, we perform the last stage of the training method to fine-tune the network. 
More specifically, we train the network for a maximum of $200$ iterations until the network estimated portfolio 
liability for the validation portfolio is within $\delta = 0.01$ relative distance of the \ac{MC} estimated 
portfolio liability of the validation portfolio. 
If the fine-tuning of the network is unable to estimate the liability of the validation portfolio 
within the defined $\delta$ relative distance, we define a new network using the set of representative 
contracts, the training portfolio and the validation portfolio and train the new network-- i.e., we do 
the complete training. We then use the new trained network in the subsequent liability estimation-- i.e., 
we use the new trained network to do the fine-tuning and portfolio liability estimation 
for subsequent $C_{s_1}^{(i)}$ values.

The idea behind the above-mentioned proposal to reduce the training time is that if two market conditions are 
very similar, the liability values of the \acp{VA} in the input portfolio under both market conditions should also 
be very close and hence the optimal network parameters (weight parameters and bias parameters) for both markets are 
likely very close to each other as well. If the optimal network parameters are indeed very close, the fine-tuning 
stage allows us to reach the optimal network parameters for the new market conditions without going through 
our computationally expensive training stage that searches for the local minimum in the whole space. 
If the fine-tuning stage fails, then we can conclude that the local minimum has changed significantly and hence a 
re-training in the whole space is required. 

To effectively exploit the closeness of market conditions to reduce the training time, we sort the $C_{s_1}^{(i)}, 1 \leq i \leq 100,$ 
values and evaluate the portfolio liability values in order. Otherwise, we might have a scenario in which consecutive 
values of $C_{s_1}^{(i)}$ represent market conditions that are not relatively close. Under such a scenario, the fine-tuning 
stages will most likely fail, requiring us to do a complete training of the neural network for each $C_{s_1}^{(i)}, 1 \leq i \leq 100$. 
The above-mentioned proposal is not as straightforward for more complex models in which the dynamics of financial markets 
is described with more than one variable. The choice of an effective strategy to exploit the closeness of market conditions 
to reduce the training time for more complex models of financial markets requires further investigation and we leave it as future work.

We compare the performance of the interpolation schemes using $6$ different realizations, 
$S_i, 1 \leq i \leq 6$, of the representative contracts, the training portfolio, and the validation portfolio. 
Table \ref{tb:rel_err1} lists the accuracy of our proposed scheme in estimating $\text{MVL}_0$, 
the $99.5\%$-quantile of $\text{MVL}_1$ ($\text{MVL}_1^{(99.5)}$), which corresponds to the $99.5\%$-quantile 
of the $\Delta$, and the \ac{SCR} value for each scenario. Accuracy is recorded as the relative error

\begin{equation}\label{eq:relative_err}
 \text{Err} = \frac{X_{NN} - X_{MC}}{|X_{MC}|}
\end{equation}
where $X_{MC}$ is the value of interest (liability or the \ac{SCR}) in the input portfolio computed by \ac{MC} simulations 
and $X_{NN}$ is the estimation of the corresponding value of interest computed by the proposed neural network method.

The results of Table \ref{tb:rel_err1} provide strong evidence that our neural network method is very accurate in its estimation 
of $\text{MVL}_0$ and $\text{MVL}_1^{(99.5)}$. The estimated liability values also result in very accurate estimation 
of the \ac{SCR}, except for scenarios $S_4$ and $S_5$. Even for scenarios $S_4$ and $S_5$, the estimated \text{SCR} values 
are well within the desired accuracy range required by insurance companies in practice. Our numerical experiments in \citep{Hejazi15-2} 
show that our proposed neural network framework has low sensitivity to the particular realization of the representative contracts, 
and the training/validation portfolio once the size of these portfolios are fixed. The results of Table \ref{tb:rel_err1} further 
corroborate our finding in \citep{Hejazi15-2} as the realization of the representative contracts, and the training/validation portfolio 
is different in each scenario.

\begin{table}[!bt]
 \centering
 \begin{tabular}{|l|l|l|l|l|l|l|}
  \hline
  \multirow{2}{*}{Value Of Interest} & \multicolumn{6}{|c|}{Relative Error (\%)}\\
  \cline{2-7} 
  & S1 & S2 & S3 & S4 & S5 & S6\\
  \hline
  \ac{SCR} & $-0.85$ & $0.69$ & $-0.81$ & $-3.58$ & $3.02$ & $1.52$\\ 
  \hline
  $\text{MVL}_0$ & $0.22$ & $0.52$ & $0.96$ & $-0.36$ & $-0.27$ & $0.70$ \\
  \hline
  $\text{MVL}_1^{(99.5)}$ & $0.43$ & $0.11$ & $0.91$ & $0.97$ & $-1.20$ & $-0.05$ \\
  \hline
 \end{tabular}
 \caption{Relative error in the estimation of the current liability value, one year liability value, and the \ac{SCR} 
 for the input portfolio.}
 \label{tb:rel_err1}
\end{table}

\begin{figure}[!bt]
\centering
\begin{subfigure}{\textwidth}
\centering
 \includegraphics[width=\linewidth, height=0.40\textheight,trim={5cm 0 5cm 1cm},clip]{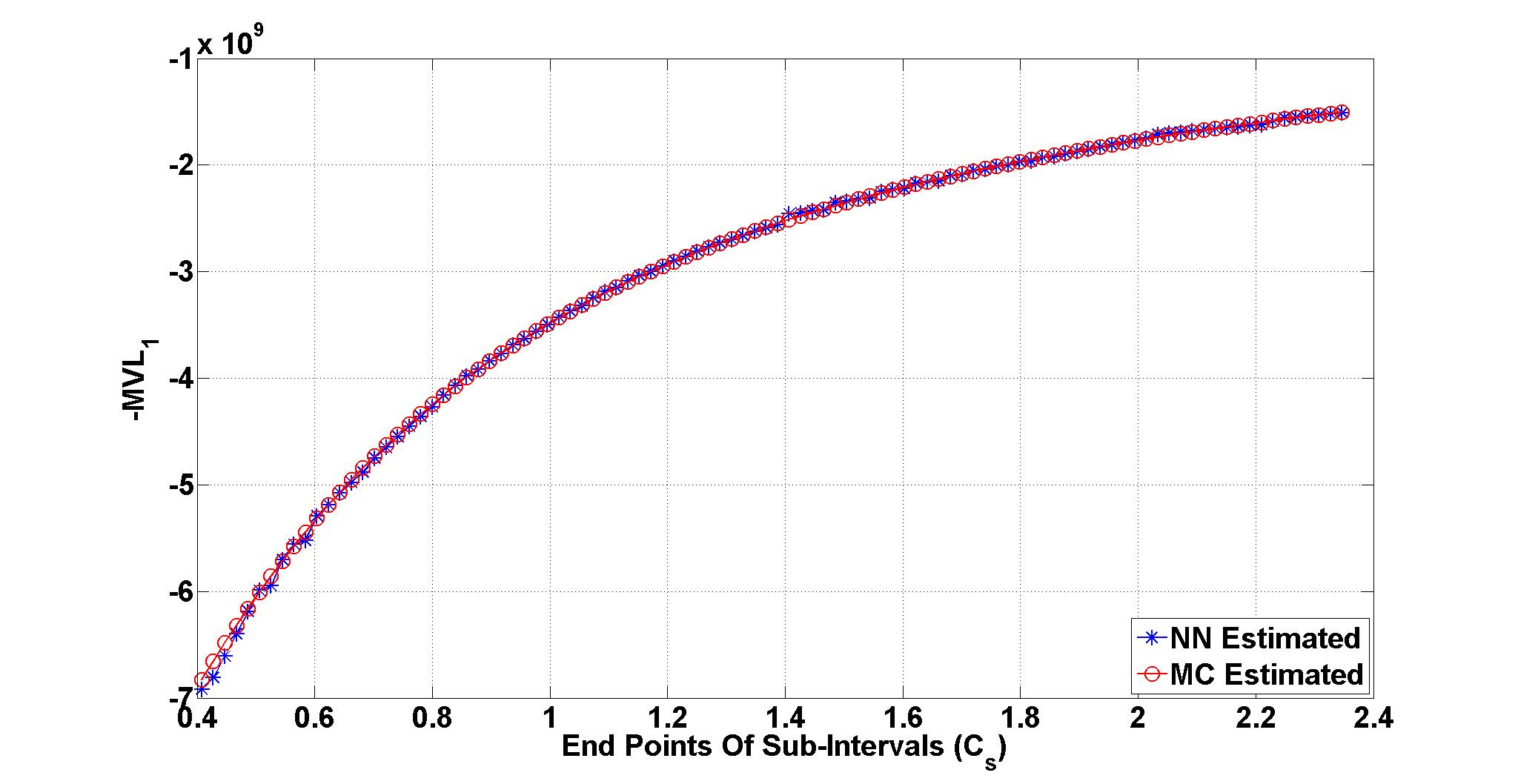}
 \caption{Estimated one-year liability values computed by the neural network framework and the \ac{MC} method.}
 \label{fig:pointwise-comp}
\end{subfigure}
\begin{subfigure}{0.47\textwidth}
 \centering
 \includegraphics[width=\linewidth, height=0.3\textheight, trim={5cm 0 5cm 0},clip]{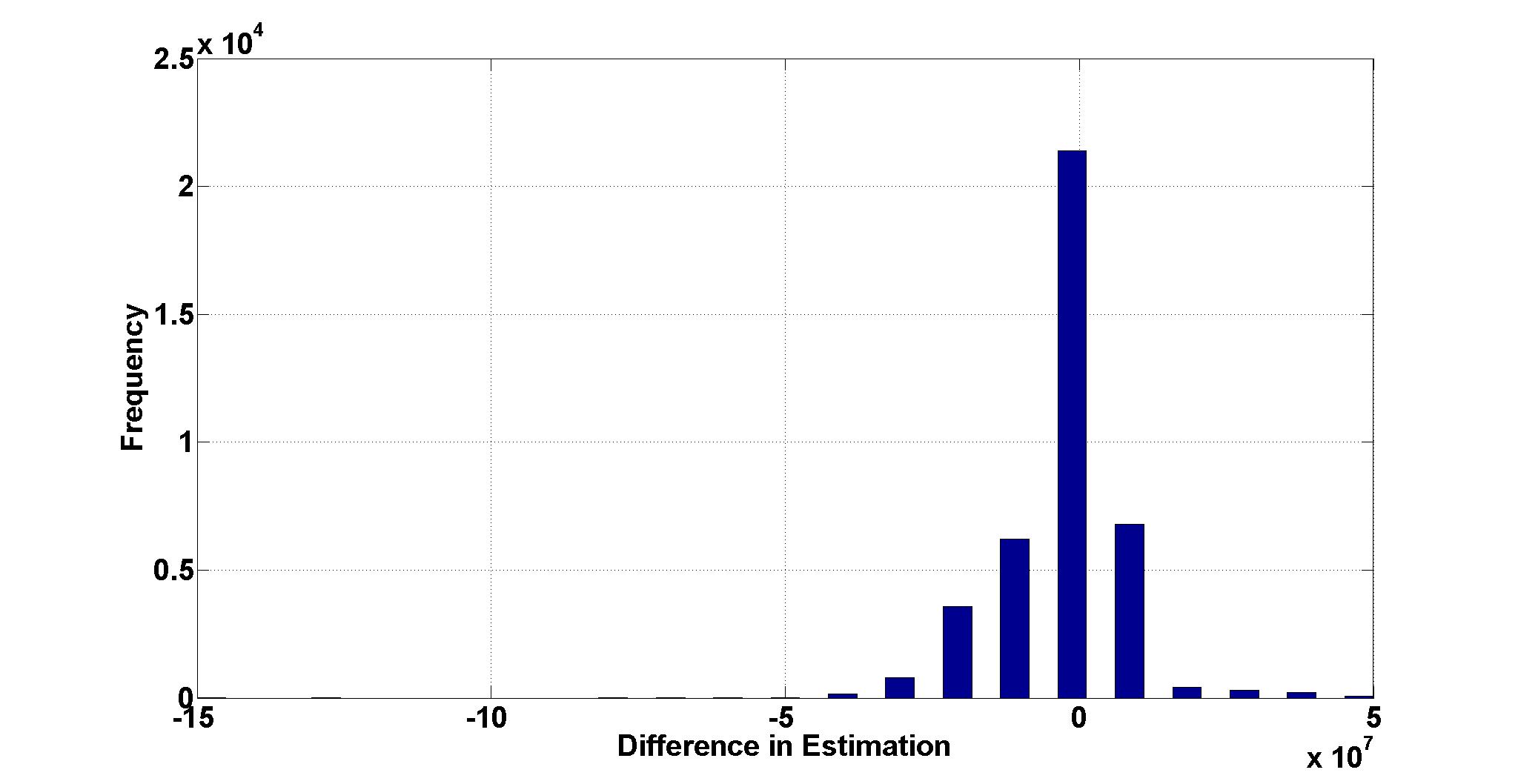}
 \subcaption{Histogram of the difference in estimation of the liability via the neural network approach and the \ac{MC} simulations at each end point of sub-intervals.}
 \label{fig:hist-abs-err}
\end{subfigure}
\begin{subfigure}{0.5\textwidth}
 \centering
 \includegraphics[width=\linewidth, height=0.3\textheight, trim={5cm 0 5cm 0},clip]{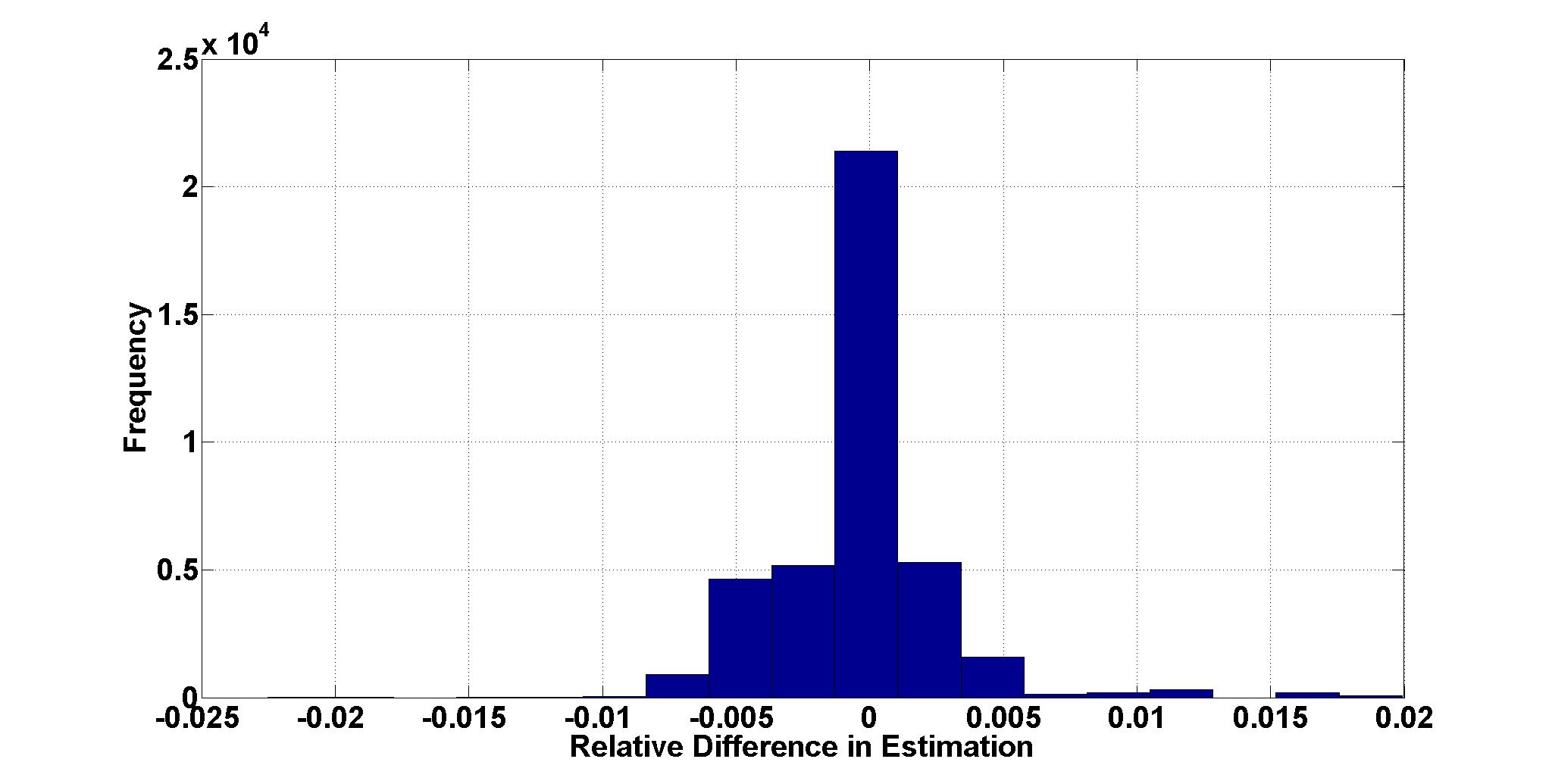}
 \subcaption{Histogram of the relative difference \eqref{eq:relative_err} in estimation of liability via the neural network approach and the \ac{MC} simulations at each end point of sub-intervals.}
 \label{fig:hist-rel-err}
\end{subfigure} 
\caption{Comparing estimation of one-year liability values of the input portfolio computed by the proposed neural network framework and the \ac{MC} method.}
\label{fig:comp-liab}
\end{figure}

The accuracy of the proposed method can be further examined by considering Figure \ref{fig:comp-liab} 
in which the estimated liability values, for the $100$ end points of the intervals, by the proposed 
neural network method are compared with their respective \ac{MC} estimations. The graphs of Figures \ref{fig:hist-abs-err} and 
\ref{fig:hist-rel-err} show that the liability estimated values 
by the proposed neural network method are very close to those of the \ac{MC} method, which demonstrates the projection capabilities of 
the neural network framework. As we discuss earlier, the smoothness of the \ac{MC} liability curve, as shown in Figure \ref{fig:comp-liab}, 
motivated us to use piecewise linear interpolation to estimate the value of the liability for points inside the sub-intervals. However, 
the estimation of this curve using the proposed neural network framework does not result in as smooth a curve as we had hoped. Therefore, we might be able 
to increase the accuracy of the proposed framework by using a non-linear curve fitting technique. Notice that, because of the particular 
simple asset structure that we use in this paper the difference between the liability values in one year's time and the corresponding $\Delta$ 
values is a constant.

As we mention in Section \ref{sec:ns_approach} and earlier in this section, in this paper, we do not address the issue of the choice 
of the interpolation scheme to estimate the $\Delta$ values of the $\text{P}^{(i)}, 1 \leq i \leq 40,000,$ paths. Because of that 
and to have a fair pointwise comparison between the proposed neural network technique and the \ac{MC} technique, we avoid using 
different interpolation schemes to estimate liabilities for the $\text{P}^{(i)}, 1 \leq i \leq 40,000,$ paths.

Table \ref{tb:sim_time1} presents the statistics on the running time of the proposed neural network approach for the nested simulation framework 
(denoted as NN in this table and elsewhere in the paper) and the nested \ac{MC} simulation framework. The results suggest a speed-up of $4-8$ times, 
depending on the scenario, and an average speed-up of $6$ times. Considering that the implementation of the neural network was 
sequential and we compared the running time of the neural network with the implementation of \ac{MC} simulations that uses parallel 
processing on 4 cores, we observe that even a simple implementation of the neural network can be highly efficient. 
As noted earlier, there is significant potential for parallelism in our neural network approach as well. Exploiting this parallelism should 
further improve the running time of the neural network. We plan to investigate this in a future paper. 
In addition, notice that we used a moderate number of \ac{MC} simulation scenarios compared to the suggested values in \citep{Bauer12}. 
As we mention earlier, an increase in the number of \ac{MC} scenarios will 
not increase the running time of our neural network significantly, because we only need \ac{MC} simulations for the representative 
contracts and for the validation/training portfolio; however, it increases the running time of the \ac{MC} simulations significantly.

\begin{table}[!bt]
 \centering
 \begin{tabular}{|l|r|r|}
  \hline
  \multirow{2}{*}{Method} & \multicolumn{2}{|c|}{Running Time}\\
  \cline{2-3} 
  & Mean & STD\\
  \hline
  MC & $49334$ & $0$\\
  \hline
  NN & $8370$ & $2465$\\ 
  \hline
 \end{tabular}
 \caption{simulation time of each method to estimate the \ac{SCR}. All times are in seconds.}
 \label{tb:sim_time1}
\end{table}

\section{Concluding Remarks}\label{sec:conclusion}
The new regulatory framework of Solvency II has been introduced by the European Union to reduce 
the risk facing insurance/re-insurance companies. An important part of the 
new regulations is the calculation of the \ac{SCR}. Because of the imprecise language used to 
describe the standards, many insurance companies are struggling to understand and implement the 
framework. 

In recent years, mathematical frameworks for calculation of the \ac{SCR} have been proposed to 
address the former issue \citep{Christiansen14, Bauer12}. Furthermore, \citep{Bauer12} 
has suggested a nested \ac{MC} simulation approach to calculate the \ac{SCR} to 
address the latter issue. The suggested \ac{MC} approach is computationally expensive, even 
for one simple insurance contract. The computational complexity of the framework stems from 
two factors: 1) the large number of outer simulation scenarios, representing different market conditions, 
required to estimate the probability distribution of the $\Delta$ and 2) the need to perform a 
computationally demanding \ac{MC} simulation for each outer simulation scenario  
to calculate the value of the liability for the insurance policy under that scenario. 

In this paper, we focus on the latter issue for a large portfolio of insurance products and propose a 
spatial interpolation approach to be used within the nested \ac{MC} simulation framework for the liability 
calculation that uses a neural network engine to interpolate 
the liability values of insurance products based on the known liability values of a small representative set 
of these products. We study the performance of the proposed approach in finding the \ac{SCR} value for a 
portfolio of \ac{VA} products. The results of our numerical experiments in Section \ref{sec:ne} corroborate the 
superior accuracy and efficiency of the sequential implementation of our proposed neural 
network approach compared with an implementation of the standard nested \ac{MC} simulation approach that uses 
parallel processing to do the \ac{MC} simulations. 

Although our method requires us to train our neural network using three small ($< 1\%$ of 
the input portfolio) portfolios that are selected uniformly at random, given an appropriate size 
for each of the small portfolios, the performance of the method has low sensitivity to the 
particular realization of these portfolios. 

Despite the superior performance of the proposed approach that uses a simple uniform 
sampling method to select the small portfolios required to train the network, 
we believe our neural network approach can be further improved by incorporating a more sophisticated 
sampling method that takes into account the distribution of the input portfolio. We intend to address 
this issue in our future research. 

The neural network that we used in our experiments was implemented sequentially. Given the structure of 
parameters/variables that define the behavior of the neural network, we can easily develop a parallel 
implementation of the neural network. A parallel implementation can significantly reduce the training time 
of the neural network and thereby the running time of the neural network framework. We plan to address 
this issue in our future work.

Although, in this paper, we do not study in depth the problem of having a large number of outer simulation scenarios,  
we proposed a possible solution via data interpolation to alleviate this problem. In our experiments, to 
reduce the simulation times and because of the smoothness of the curve that describes the $\Delta$ values, as a 
first simple choice, we suggest using piecewise-linear interpolation to approximate the $\Delta$ values. However, 
our experimental results suggest that using a better interpolation method can increase the accuracy of our proposed 
neural network approach for nested simulations. We intend to study the choice of the interpolation scheme in our future research.

In this introductory paper on our neural network approach, we chose to focus on a simple model of financial markets 
to make our analysis more tractable. However, as we discuss in Sections \ref{sec:ns_approach} and \ref{sec:ne}, 
the approach can be extended to incorporate more complex models of financial markets. To achieve the best outcome 
with our neural network approach, we need to study an effective strategy to exploit the closeness of the sample 
points representing various states of the financial markets to reduce the training time of the neural network. 
We intend to study this in our future work.

In this paper, we chose to study the performance of our neural network approach on estimation of liabilities; however, the 
application of our proposed approach is much more general than that. In particular, one can 
change the type of the insurance product and the approach used to value individual insurance products 
and incorporate them within our framework to estimate the value of a large portfolio of the aforementioned 
insurance products. 

\section{Acknowledgements}
This research was supported in part by the Natural Sciences and Engineering Research Council of Canada (NSERC).

\bibliographystyle{elsarticle-num-names}
\bibliography{CAM-Paper3}







\end{document}